\pdfoutput=1
\documentclass[12pt,a4paper]{article}
\usepackage{ifthen} % for conditional statements
\newboolean{pdflatex}
\setboolean{pdflatex}{true} % False for eps figures 

\newboolean{articletitles}
\setboolean{articletitles}{true} % False removes titles in references

\newboolean{uprightparticles}
\setboolean{uprightparticles}{false} %True for upright particle symbols

\def\paperauthors{LHCb collaboration} % Leave as is for PAPER, CONF and FIGURE
\def\paperasciititle{Measurement of $CP$ asymmetry in $D^0 \rightarrow K_S^0K_S^0$ decays} % Set ASCII title here !! MAKE sure it's only ASCII characters !! 
\def\papertitle{Measurement of \CP asymmetry\\ in \Dz\to\ \KS\KS decays} % Latex formatted title
\def\paperkeywords{{High Energy Physics}, {LHCb}} % Comma separated list
\def\papercopyright{\the\year\ CERN for the benefit of the LHCb collaboration} % new since 9/Apr/2018
\def\paperlicence{CC BY 4.0 licence}
\def\paperlicenceurl{https://creativecommons.org/licenses/by/4.0/}

\usepackage[top=1in, bottom=1.25in, left=1in, right=1in]{geometry}

\usepackage{microtype}
\usepackage{lineno}  % for line numbering during review
\usepackage{xspace} % To avoid problems with missing or double spaces after
\usepackage{caption} %these three command get the figure and table captions automatically small

\usepackage{graphicx}  % to include figures (can also use other packages)
\usepackage{color}
\usepackage{colortbl}
\graphicspath{{./figs/}} % Make Latex search fig subdir for figures
\usepackage{amsmath} % Adds a large collection of math symbols
\usepackage{amssymb}
\usepackage{amsfonts}
\usepackage{upgreek} % Adds in support for greek letters in roman typeset

\newcommand*\patchAmsMathEnvironmentForLineno[1]{%
\expandafter\let\csname old#1\expandafter\endcsname\csname #1\endcsname
\expandafter\let\csname oldend#1\expandafter\endcsname\csname
end#1\endcsname
 \renewenvironment{#1}%
   {\linenomath\csname old#1\endcsname}%
   {\csname oldend#1\endcsname\endlinenomath}%
}
\newcommand*\patchBothAmsMathEnvironmentsForLineno[1]{%
  \patchAmsMathEnvironmentForLineno{#1}%
  \patchAmsMathEnvironmentForLineno{#1*}%
}
\AtBeginDocument{%
\patchBothAmsMathEnvironmentsForLineno{equation}%
\patchBothAmsMathEnvironmentsForLineno{align}%
\patchBothAmsMathEnvironmentsForLineno{flalign}%
\patchBothAmsMathEnvironmentsForLineno{alignat}%
\patchBothAmsMathEnvironmentsForLineno{gather}%
\patchBothAmsMathEnvironmentsForLineno{multline}%
\patchBothAmsMathEnvironmentsForLineno{eqnarray}%
}

\usepackage{hyperxmp}

\usepackage[pdftex,
            pdfauthor={\paperauthors},
            pdftitle={\paperasciititle},
            pdfkeywords={\paperkeywords},
            pdfcopyright={Copyright (C) \papercopyright},
            pdflicenseurl={\paperlicenceurl}]{hyperref}
\usepackage[colorinlistoftodos,textsize=scriptsize]{todonotes}

\usepackage[bottom,flushmargin,hang,multiple]{footmisc}

\usepackage[all]{hypcap} % Internal hyperlinks to floats.
\usepackage{booktabs}
\usepackage{xspace} 
\usepackage{upgreek}

\def\lhcb   {\mbox{LHCb}\xspace}

\def\MagUp {\mbox{\em Mag\kern -0.05em Up}\xspace}
\def\MagDown {\mbox{\em MagDown}\xspace}

\ifthenelse{\boolean{uprightparticles}}%
{

 \def\Ppi         {\ensuremath{\uppi}\xspace}

 \def\PDelta      {\ensuremath{\Delta}\xspace}                 
 \def\PXi         {\ensuremath{\Xi}\xspace}                 
 \def\PLambda     {\ensuremath{\Lambda}\xspace}                 
 \def\PSigma      {\ensuremath{\Sigma}\xspace}                 
 \def\POmega      {\ensuremath{\Omega}\xspace}                 
 \def\PUpsilon    {\ensuremath{\Upsilon}\xspace}

 \def\PB      {\ensuremath{\mathrm{B}}\xspace}                 
                  
 \def\PD      {\ensuremath{\mathrm{D}}\xspace}

 \def\PK      {\ensuremath{\mathrm{K}}\xspace}

 \def\Pb      {\ensuremath{\mathrm{b}}\xspace}                 
 \def\Pc      {\ensuremath{\mathrm{c}}\xspace}                 
 \def\Pd      {\ensuremath{\mathrm{d}}\xspace}

 \def\Pi      {\ensuremath{\mathrm{i}}\xspace}

 \def\Pp      {\ensuremath{\mathrm{p}}\xspace}

 \def\Ps      {\ensuremath{\mathrm{s}}\xspace}                 
 \def\Pt      {\ensuremath{\mathrm{t}}\xspace}                 
 \def\Pu      {\ensuremath{\mathrm{u}}\xspace}

 \def\thebaroffset{0.0em}
}
{

 \def\Ppi         {\ensuremath{\pi}\xspace}

 \mathchardef\PDelta="7101
 \mathchardef\PXi="7104
 \mathchardef\PLambda="7103
 \mathchardef\PSigma="7106
 \mathchardef\POmega="710A
 \mathchardef\PUpsilon="7107
                  
 \def\PB      {\ensuremath{B}\xspace}                 
                  
 \def\PD      {\ensuremath{D}\xspace}

 \def\PK      {\ensuremath{K}\xspace}

 \def\Pb      {\ensuremath{b}\xspace}                 
 \def\Pc      {\ensuremath{c}\xspace}                 
 \def\Pd      {\ensuremath{d}\xspace}

 \def\Pi      {\ensuremath{i}\xspace}

 \def\Pp      {\ensuremath{p}\xspace}

 \def\Ps      {\ensuremath{s}\xspace}                 
 \def\Pt      {\ensuremath{t}\xspace}                 
 \def\Pu      {\ensuremath{u}\xspace}

 \def\thebaroffset{0.18em}
}
\newcommand{\offsetoverline}[2][\thebaroffset]{\kern #1\overline{\kern -#1 #2}}%

\makeatletter
\ifcase \@ptsize \relax% 10pt
  \newcommand{\miniscule}{\@setfontsize\miniscule{4}{5}}% \tiny: 5/6
\or% 11pt
  \newcommand{\miniscule}{\@setfontsize\miniscule{5}{6}}% \tiny: 6/7
\or% 12pt
  \newcommand{\miniscule}{\@setfontsize\miniscule{5}{6}}% \tiny: 6/7
\fi
\makeatother

\DeclareRobustCommand{\optbar}[1]{\shortstack{{\miniscule (\rule[.5ex]{1.25em}{.18mm})}
  \\ [-.7ex] $#1$}}

\def\uquark    {{\ensuremath{\Pu}}\xspace}

\def\dquark    {{\ensuremath{\Pd}}\xspace}

\def\squark    {{\ensuremath{\Ps}}\xspace}

\def\cquark    {{\ensuremath{\Pc}}\xspace}

\def\bquark    {{\ensuremath{\Pb}}\xspace}

\def\tquark    {{\ensuremath{\Pt}}\xspace}

\def\pion   {{\ensuremath{\Ppi}}\xspace}
\def\piz    {{\ensuremath{\pion^0}}\xspace}
\def\pip    {{\ensuremath{\pion^+}}\xspace}
\def\pim    {{\ensuremath{\pion^-}}\xspace}

\def\kaon    {{\ensuremath{\PK}}\xspace}
\def\KorKbar {\kern \thebaroffset\optbar{\kern -\thebaroffset \PK}{}\xspace}

\def\Kp      {{\ensuremath{\kaon^+}}\xspace}
\def\Km      {{\ensuremath{\kaon^-}}\xspace}

\def\KS      {{\ensuremath{\kaon^0_{\mathrm{S}}}}\xspace}

\def\Dbar    {{\ensuremath{\offsetoverline{\PD}}}\xspace}
\def\D       {{\ensuremath{\PD}}\xspace}

\def\DorDbar {\kern \thebaroffset\optbar{\kern -\thebaroffset \PD}\xspace}
\def\Dz      {{\ensuremath{\D^0}}\xspace}
\def\Dzb     {{\ensuremath{\Dbar{}^0}}\xspace}
\def\Dp      {{\ensuremath{\D^+}}\xspace}
\def\Dm      {{\ensuremath{\D^-}}\xspace}

\def\DpDm    {\ensuremath{\Dp {\kern -0.16em \Dm}}\xspace}
\def\Dstar   {{\ensuremath{\D^*}}\xspace}

\def\Dstarp  {{\ensuremath{\D^{*+}}}\xspace}

\def\Dstarpm {{\ensuremath{\D^{*\pm}}}\xspace}

\def\Dsp     {{\ensuremath{\D^+_\squark}}\xspace}

\def\B       {{\ensuremath{\PB}}\xspace}

\def\BorBbar {\kern \thebaroffset\optbar{\kern -\thebaroffset \PB}\xspace}

\def\Bd      {{\ensuremath{\B^0}}\xspace}

\def\BdorBdbar {\kern \thebaroffset\optbar{\kern -\thebaroffset \Bd}\xspace}

\def\Bs      {{\ensuremath{\B^0_\squark}}\xspace}

\def\BsorBsbar {\kern \thebaroffset\optbar{\kern -\thebaroffset \Bs}\xspace}

\def\Y#1S{\ensuremath{\PUpsilon{(#1S)}}\xspace}

\def\proton      {{\ensuremath{\Pp}}\xspace}

\def\LorLbar     {\kern \thebaroffset\optbar{\kern -\thebaroffset \PLambda}\xspace}

\newcommand{\decay}[2]{\ensuremath{#1\!\to #2}\xspace} 

\def\to                 {\ensuremath{\rightarrow}\xspace}

\def\eps   {{\ensuremath{\varepsilon}}\xspace}

\def\CP                {{\ensuremath{C\!P}}\xspace}
\def\CPV               {{\ensuremath{C\!PV}}\xspace}

\newcommand{\Delm}{{\mbox{$\Delta m $}}\xspace}
\newcommand{\ACP}{{\ensuremath{{\mathcal{A}}^{\CP}}}\xspace}

\def\AT#1     {\ensuremath{A_{\mathrm{T}}^{#1}}\xspace}           % 2

\def\C#1      {\ensuremath{\mathcal{C}_{#1}}\xspace}                       % 9
\def\Cp#1     {\ensuremath{\mathcal{C}_{#1}^{'}}\xspace}                    % 7
\def\Ceff#1   {\ensuremath{\mathcal{C}_{#1}^{\mathrm{(eff)}}}\xspace}        % 9  
\def\Cpeff#1  {\ensuremath{\mathcal{C}_{#1}^{'\mathrm{(eff)}}}\xspace}       % 7
\def\Ope#1    {\ensuremath{\mathcal{O}_{#1}}\xspace}                       % 2
\def\Opep#1   {\ensuremath{\mathcal{O}_{#1}^{'}}\xspace}                    % 7

\newcommand{\aunit}[1]{\ensuremath{\text{\,#1}}}       
\newcommand{\tev}{\aunit{Te\kern -0.1em V}\xspace}
\newcommand{\gev}{\aunit{Ge\kern -0.1em V}\xspace}
\newcommand{\mev}{\aunit{Me\kern -0.1em V}\xspace}
\newcommand{\kev}{\aunit{ke\kern -0.1em V}\xspace}
\newcommand{\ev}{\aunit{e\kern -0.1em V}\xspace}
 
\newcommand{\mevc}{\ensuremath{\aunit{Me\kern -0.1em V\!/}c}\xspace}
\newcommand{\gevc}{\ensuremath{\aunit{Ge\kern -0.1em V\!/}c}\xspace}
\newcommand{\mevcc}{\ensuremath{\aunit{Me\kern -0.1em V\!/}c^2}\xspace}
\newcommand{\gevcc}{\ensuremath{\aunit{Ge\kern -0.1em V\!/}c^2}\xspace}
\def\fb   {\ensuremath{\aunit{fb}}\xspace}
\def\invfb   {\ensuremath{\fb^{-1}}\xspace}

\newcommand{\chisq}{\ensuremath{\chi^2}\xspace}

\def\gsim{{~\raise.15em\hbox{$>$}\kern-.85em
          \lower.35em\hbox{$\sim$}~}\xspace}
\def\lsim{{~\raise.15em\hbox{$<$}\kern-.85em
          \lower.35em\hbox{$\sim$}~}\xspace}

\def\tell1  {TELL1\xspace}
\def\ukl1   {UKL1\xspace}

\usepackage{cite} % Allows for ranges in citations
\usepackage{mciteplus}
\usepackage{longtable} % only for template; not usually to be used in PAPERs

\begin{document}

%%%%%%%%%%%%%%%%%%%%%%%%%
%%%%% Title     %%%%%%%%%
%%%%%%%%%%%%%%%%%%%%%%%%%
\renewcommand{\thefootnote}{\fnsymbol{footnote}}
\setcounter{footnote}{1}

% %%%%%%% CHOOSE TITLE PAGE--------
%\onecolumn
%\input{title-LHCb-INT}
%\input{title-LHCb-ANA}
%\input{title-LHCb-CONF}
%\input{title-LHCb-FIGURE}
% ===============================================================================
% Purpose: LHCb-PAPER journal paper title page template
% Author: 
% Created on: 2010-09-25
% ===============================================================================

%%%%%%%%%%%%%%%%%%%%%%%%%
%%%%%  TITLE PAGE  %%%%%%
%%%%%%%%%%%%%%%%%%%%%%%%%
\begin{titlepage}
\pagenumbering{roman}

% Header ---------------------------------------------------
\vspace*{-1.5cm}
\centerline{\large EUROPEAN ORGANIZATION FOR NUCLEAR RESEARCH (CERN)}
\vspace*{1.5cm}
\noindent
\begin{tabular*}{\linewidth}{lc@{\extracolsep{\fill}}r@{\extracolsep{0pt}}}
\ifthenelse{\boolean{pdflatex}}% Logo format choice
{\vspace*{-1.5cm}\mbox{\!\!\!\includegraphics[width=.14\textwidth]{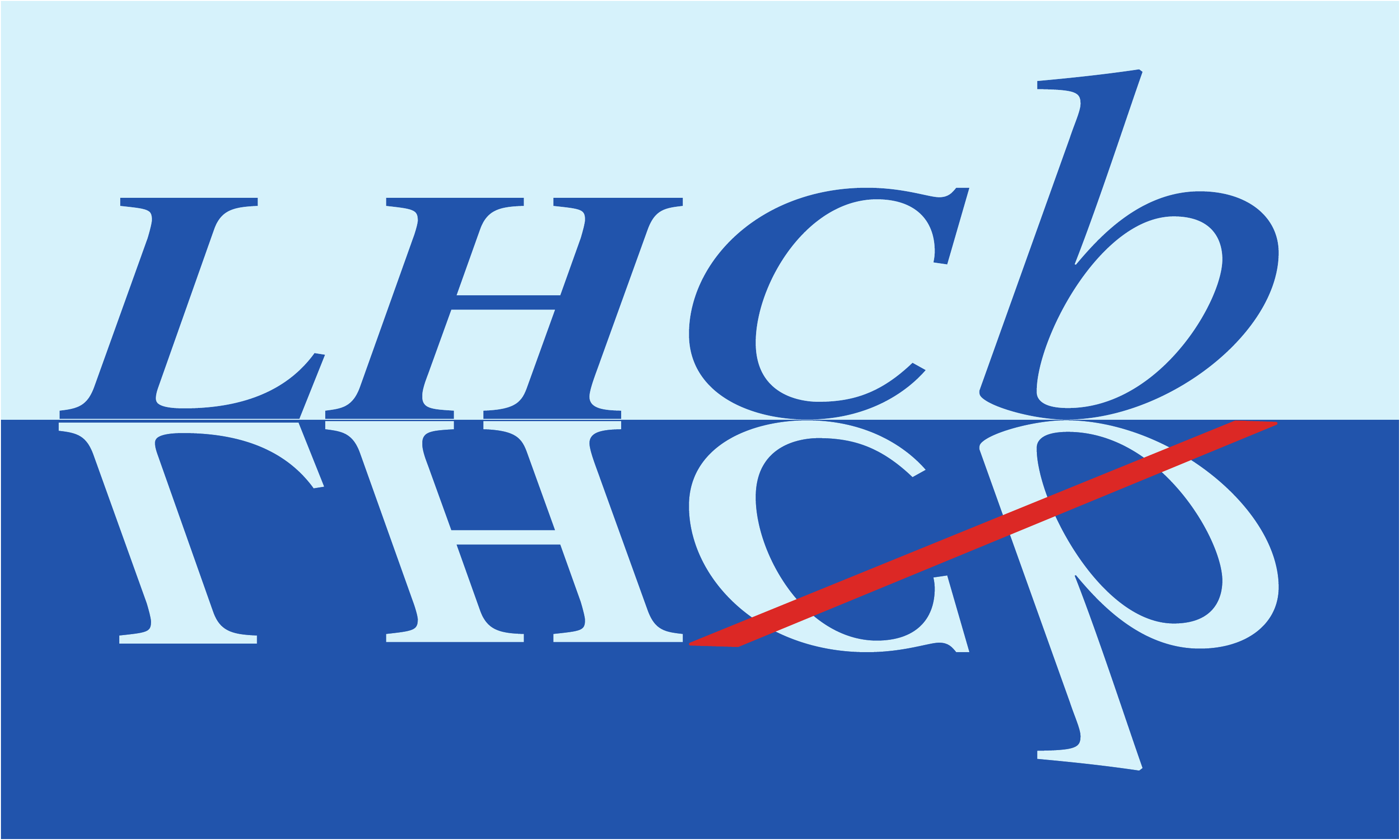}} & &}%
{\vspace*{-1.2cm}\mbox{\!\!\!\includegraphics[width=.12\textwidth]{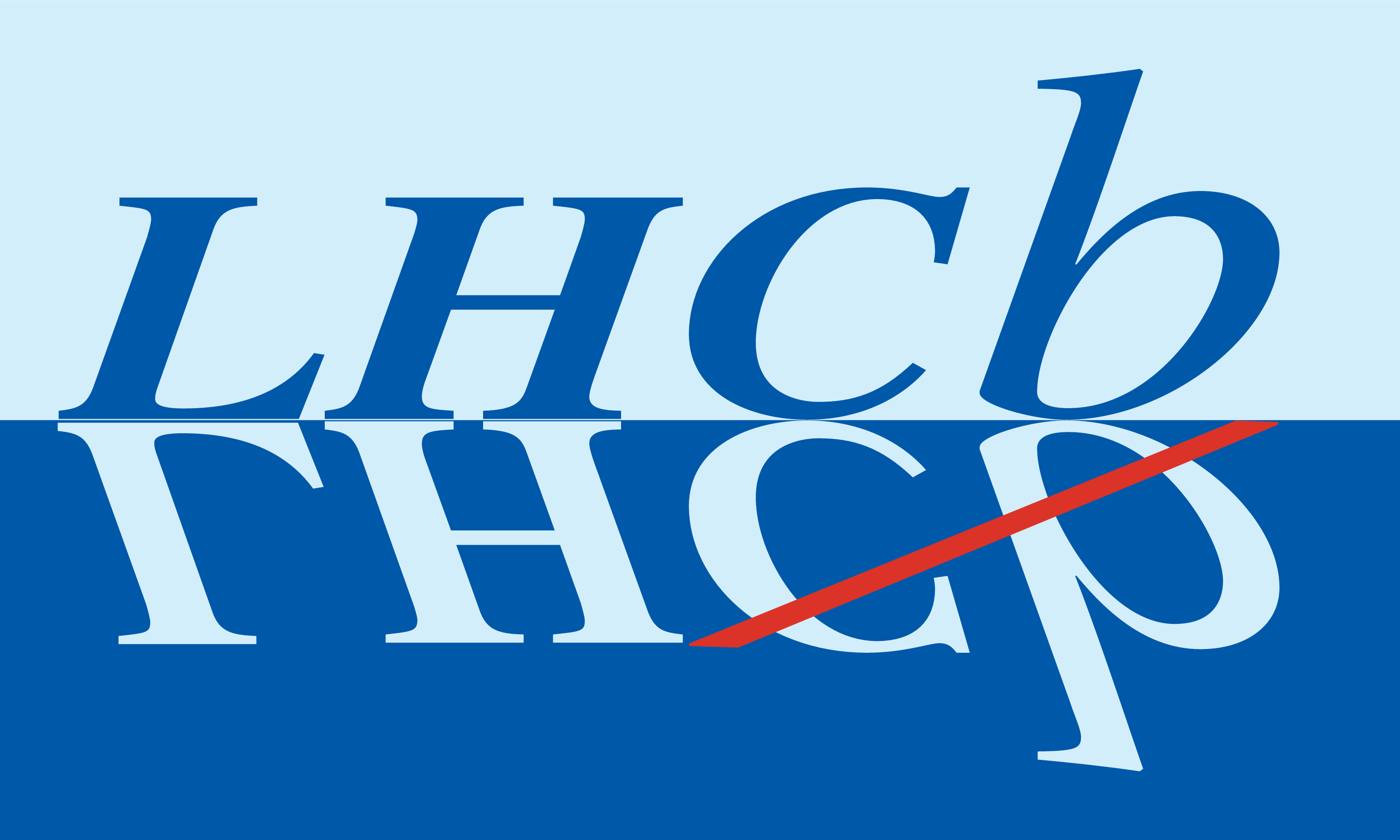}} & &}%
\\
 & & CERN-EP-2021-058 \\  % ID 
 & & LHCb-PAPER-2020-047 \\  % ID 
 & & September 15, 2021 \\ % Date - Can also hardwire e.g.: 23 March 2010
 & & \\
% not in paper \hline
\end{tabular*}

\vspace*{4.0cm}

% Title --------------------------------------------------
{\normalfont\bfseries\boldmath\huge
\begin{center}
% DO NOT EDIT HERE. Instead edit macro in main.tex to keep metadata correct
  \papertitle 
\end{center}
}

\vspace*{2.0cm}

% Authors -------------------------------------------------
\begin{center}
%In the footnote, replace 'paper' by 'Letter' in case of submission to PRL or PLB 
% Edit macro in main.tex to keep metadata correct
\paperauthors\footnote{Authors are listed at the end of this Letter.}
\end{center}

\vspace{\fill}

% Abstract -----------------------------------------------
\begin{abstract}
  \noindent
A measurement of the \CP asymmetry in $\Dz \to \KS \KS$ decays is reported, based on a data sample of proton-proton collisions collected by the LHCb experiment from 2015 to 2018, corresponding to an integrated luminosity of $6\invfb$. The flavor of the \Dz candidate is determined using the charge of the \Dstarpm meson, from which the decay is required to originate.  The $\Dz \to \Kp \Km$ decay is used as a calibration channel. The  time-integrated \CP asymmetry for the $\Dz\to \KS \KS$ mode is measured to be:
$$\ACP(\Dz\to \KS \KS) = (-3.1\pm 1.2\pm 0.4 \pm 0.2)\%, $$
where the first uncertainty is statistical, the second is systematic, and the third is due to the uncertainty on the \CP asymmetry of the calibration channel.
This is the most precise determination of this quantity to date.
  
\end{abstract}

\vspace*{2.0cm}

\begin{center}
Published as a Letter in
%  JHEP /
%  Phys.~Rev.~D /
  Phys.~Rev.~D 104, 031102 (2021)
%  Phys.~Lett.~B /
%  Eur.~Phys.~J.~C /
  %  Nucl.~Phys.~B /
%  Chin.~Phys.~C /
%  Nature~Physics /
%  sciPost~Physics /
%  J. Instr. /
%  Instruments 
\end{center}

\vspace{\fill}

{\footnotesize 
% Edit macro in main.tex to keep metadata correct
\centerline{\copyright~\papercopyright. \href{\paperlicenceurl}{\paperlicence}.}}
\vspace*{2mm}

\end{titlepage}

%%%%%%%%%%%%%%%%%%%%%%%%%%%%%%%%
%%%%%  EOD OF TITLE PAGE  %%%%%%
%%%%%%%%%%%%%%%%%%%%%%%%%%%%%%%%

%  empty page follows the title page ----
\newpage
\setcounter{page}{2}
\mbox{~}

%\twocolumn
% %%%%%%%%%%%%% ---------

\renewcommand{\thefootnote}{\arabic{footnote}}
\setcounter{footnote}{0}

%%%%%%%%%%%%%%%%%%%%%%%%%%%%%%%%
%%%%%  Table of Content   %%%%%%
%%%%%%%%%%%%%%%%%%%%%%%%%%%%%%%%
%%%% Uncomment if desired
%\tableofcontents
\cleardoublepage

%%%%%%%%%%%%%%%%%%%%%%%%%
%%%%% Main text %%%%%%%%%
%%%%%%%%%%%%%%%%%%%%%%%%%

\pagestyle{plain} % restore page numbers for the main text
\setcounter{page}{1}
\pagenumbering{arabic}

%% Uncomment during review phase. 
%% Comment before a final submission.
%\linenumbers

%% This is the main body
%% It is useful to have a single file so comemnts are not missed in overleaf.
The existence of charge-parity violation (\CPV) effects in  charm hadrons has recently been established \cite{LHCb-PAPER-2019-006}, which constitutes the only evidence of \CPV in the up-quark (\uquark, \cquark or \tquark) sector. However, current experimental evidence is limited to a single observable, $\Delta \ACP$, the difference between the \CP asymmetry in \decay{\Dz}{\Kp \Km} and $\mbox{\decay{\Dz}{\pip \pim}}$ decays. Given the uncertainties in the theoretical predictions on \CPV in the charm-quark sector within the Standard Model (SM), it is currently not possible to reach a definitive conclusion about their ability to explain the data~\cite{Chala:2019fdb, Grossman:2019xcj}.  
Further measurements in charm-hadron decays are crucial to shed light on \CPV phenomenology. 
This could involve dynamics beyond the SM, which is not constrained to be the same as in the down-quark (\dquark, \squark or \bquark) sector
and could enter the amplitudes via loop contributions, affecting observables in a detectable way.

Among many charm-hadron decay modes in which \CPV could manifest, the \mbox{\decay{\Dz}{\KS \KS}} mode is a promising target because of the expected size of the effect.
Its \CP asymmetry is defined by 
\begin{equation}\label{eq:acp_def}
\ACP(\KS\KS) = \dfrac{\Gamma (\decay{\Dz}{\KS\KS}) - \Gamma (\decay{\Dzb}{\KS \KS})}{\Gamma (\decay{\Dz}{\KS\KS}) + \Gamma (\decay{\Dzb}{\KS \KS})},
\end{equation}
where $\Gamma$ is the decay width of the \Dz (\Dzb) meson,  and it can be larger than in other channels, up to the percent level in the SM~\cite{Nierste:2015zra, Li:2012cfa, Cheng:2019ggx, Buccella:2019kpn, Brod:2011re}.   
In fact, only amplitudes proceeding via loop-suppressed and tree-level exchange diagrams, which vanish in the flavor--SU(3) limit, contribute to this decay, and they are similar in size. 
Their interference could therefore result in a detectable \CP asymmetry. In addition, the \CP asymmetry in the $\decay{\Dz}{\KS \KS}$ decay is sensitive to a different mix of amplitudes compared to $\decay{\Dz}{\Kp \Km}$ and $\decay{\Dz}{\pip \pim}$ decays. 
Therefore, measuring \ACP(\KS\KS) provides independent information which can help to elucidate the mechanisms of \CPV in charm hadron decays. 

The current world average for the time-integrated \CP asymmetry is \mbox{$\ACP (\KS\KS) = (0.4\pm 1.4)\%$}\cite{PDG2020}, the precision of which is still insufficient for detecting an effect.
In this work, a new measurement of this quantity, performed with proton-proton ($pp$) collisions collected from 2015 to 2018 (Run 2) by the LHCb experiment at the LHC at CERN, is reported. Data collected at a center-of-mass energy of 13 TeV, corresponding to an integrated luminosity of $6\,\invfb$, are used.
The subsample of data collected during 2015 and 2016 has been analyzed previously \cite{LHCb-PAPER-2018-012}. In the present work a number of improvements in the analysis leads to a more efficient selection and a sizeable improvement in sensitivity. 

The measurement of $\ACP(\KS\KS)$  requires knowledge of the \Dz flavor at production.
A sample of flavor-tagged $\decay{\Dz}{\KS \KS}$ decays is obtained by selecting only \Dz mesons that originate from $\Dstarp \to \Dz \pip$ decays.\footnote{Inclusion of the charge-conjugate process is implied throughout this document unless explicitly specified.} 
The charge of the pion (tagging pion) in this decay identifies the flavor of the accompanying \Dz meson. While \Dz oscillations may cause some of them to change flavor before decaying, this is a small effect in comparison to the resolution of the current measurement and is not considered further.

The decay widths $\Gamma$ in Eq.~\ref{eq:acp_def} are related to the number of observed candidates $N$ by 
\begin{equation}
N(\DorDbar^{\rm 0} \to \KS \KS) \propto \sigma (\Dstarpm) \eps^{\pm} \Gamma (\DorDbar^{\rm 0} \to \KS \KS),    
\end{equation}
where $\sigma(\Dstarpm)$ are the production cross sections and $ \eps^{\pm}$  the detection probabilities for \Dstarpm decays. 
Both factors are charge asymmetric, due to the \Dstarpm production asymmetry arising from the hadronization of charm quarks in $pp$ collisions, and to asymmetries in the geometry and response of the detector.
Because of the charge-symmetric final state of the \Dz decay, the only  source of detection asymmetry comes from the tagging pion.  
All these quantities are calibrated using a large sample of $\decay{\Dz}{\Kp \Km}$ decays, for which \ACP is known with much greater precision than in the $\decay{\Dz}{\KS\KS}$ decay mode~\cite{LHCb-PAPER-2016-035}.

The LHCb detector is a single-arm forward spectrometer designed for the study of particles containing $b$ or $c$ quarks, as described in detail in Refs.~\cite{LHCb-DP-2008-001,LHCb-DP-2014-002}. The LHCb tracking system exploits a dipole magnet to measure the momentum of charged particles, and it 
consists of a silicon-strip vertex detector surrounding the $pp$ interaction region~\cite{LHCb-DP-2014-001} and tracking stations placed upstream and downstream of the magnet.
The magnetic-field polarity is reversed periodically during data taking, alternating between pointing upwards (\MagUp) and downwards (\MagDown), to mitigate the differences of reconstruction efficiencies of particles with opposite charges. 
The online event selection is performed by a trigger~\cite{LHCb-DP-2012-004}, 
which consists of a hardware stage, based on information from the calorimeter and muon systems, followed by a software stage, which applies a full event reconstruction and selection in two steps.
The simulated data used in this analysis are produced using the software described in Refs.~\cite{Sjostrand:2006za,LHCb-PROC-2010-056,Lange:2001uf,Allison:2006ve,LHCb-PROC-2011-006}.
 
For an event to be considered in this analysis, the hardware-trigger decision is required to have been based either on the transverse energy deposited in the hadronic calorimeter by a particle associated with the $\Dz$ candidate, or on signatures not associated with the $\Dstarp$ candidate, such as a muon with high transverse momentum, or a large transverse-energy deposit in the electromagnetic or hadronic calorimeter from the rest of the event.
Similarly, the first software stage decision is required to have been based either on the presence of a single track associated with the \Dz decay, with sufficient transverse momentum and impact parameter relative to the primary \proton\proton collision vertex (PV), or on track-related signatures independent of the $\Dstarp$ candidate. 
This ensures that the trigger decision is independent of the tagging pion, preventing a possible charge-dependent bias which would be difficult to measure and correct accurately.

At the second software stage, a complete reconstruction of the decay chain is performed, requiring a pair of \KS mesons compatible with the decay of a \Dz particle, and the presence of an additional pion, compatible with originating from a \decay{\Dstarp}{\Dz \pip} decay. Candidate \decay{\KS}{\pip\pim} decays are classified in two categories: those in which the decay occurred early enough for the pions to be reconstructed in the vertex detector (``long'', indicated in the following by an ``L'' label); and those decaying later, such that segments of pion tracks can only be formed in the  silicon-strip tracker upstream of the magnet, and the tracking stations after the magnet (``downstream'', indicated in the following by a ``D'' label). 
The geometric acceptance of the \lhcb detector for typical \KS momenta is about a factor of 2 larger for D-type versus L-type decays, but the mass, momentum and vertex resolution of the latter category are better than those of the former. In addition, downstream tracks are only reconstructed in the second-stage software trigger, causing a lower efficiency for D-type decays, as can be deduced from Table~\ref{tab:res}. 

Candidates are separated into three categories: LL, LD, or DD, according to the types of the two \KS mesons associated to the \Dz meson. These are analyzed separately, and the results are combined only at the end. To prevent any experimenters' bias, the flavor of \Dz candidates was not examined until the analysis was finalized.

Some of the \Dstarp mesons are not promptly produced in the primary \proton \proton interaction, but rather come from the decay of a beauty-flavored hadron. These secondary decays are affected by different production asymmetries with respect to prompt decays, which is estimated at the 1\% level. 
No attempt is made in this analysis to reject secondary decays, as selection requirements based on pointing observables are less effective in \Dz mesons reconstructed with displaced vertices, such as \decay{\Dz}{\KS \KS}, compared to other charm hadron decays. 
The spurious asymmetry introduced by this contribution is instead canceled by use of a dedicated calibration sample.
For this purpose, a sample of $\decay{\Dz}{\Kp \Km}$ decays produced in \Dstarp decays is selected in a way to ensure it contains prompt and secondary decays in the same proportions as the $\decay{\Dz}{\KS \KS}$ sample. 
Its selection, starting from the first trigger level, refrains from any requirements for which the effect on the secondary-to-prompt ratio cannot reliably reproduce that in the signal sample. This requires, among other things, avoiding selections based on observables sensitive to the location of the \Dstar or \Dz vertices, which have different resolutions in the two samples~\cite{tuci}.

The \decay{\Dz}{\KS \KS} signal peaks in the difference of invariant masses $\Delm = m(\Dstarp) - m(\Dz)$, where $m(\Dz)$ is the invariant mass of the \decay{\Dz}{\KS \KS} candidate and $m(\Dstarp)$ is the invariant mass of the \decay{\Dstarp}{\Dz \pip} candidate. This sample may be contaminated by other physics processes. Amongst them, partially reconstructed \Dz  decays coming from \Dstarp mesons, such as $\decay{\Dz}{\KS \KS} \piz$, can also peak in $\Delm$. Some non-\Dz decays can also contaminate the sample, the main contributor being $\decay{\Dsp}{\KS \KS \pip}$ decays, where the $\KS \KS$ pair is incorrectly associated with a $\Dz$ decay. These background components are effectively suppressed by selecting only candidates with $m(\Dz)$ within $\pm 20(30)\mevcc$ of the known \Dz mass\cite{PDG2020} in LL/LD (DD) samples.

A background source which is difficult to eliminate completely is the abundant $\mbox{\Dz \to \KS \pip \pim}$ decay, with the $\pip \pim$ pair mimicking a $\KS$. After a loose selection based on the flight distance of the \KS, it is
statistically separated from the signal through a simultaneous fit of three observables: the $\Delm$ and the invariant masses, $m(\KS)$, of the two $\KS$ candidates. 

To maximize the sensitivity of the measurement, the LL and LD subsamples are further split according to whether the \Dstarp candidate is compatible or incompatible with having originated from the~PV. Compatibility is defined by a fixed threshold on the goodness of fit of the complete decay chain~\cite{Hulsbergen:2005pu}, when constrained to be originating from the PV. For PV-compatible candidates, independent of their true nature, a factor of 2 better mass resolution is obtained by constraining the origin vertex to coincide with the PV. 
Therefore, separate analyses of the two subsamples, followed by the combination of the results, provides optimal exploitation of the available data. 
This is not done for the DD sample due to its limited signal yield.
While this requirement might in principle have slightly different effects on the secondary fraction of signal and calibration sample, differences are negligible, as it has been explicitly verified on data~\cite{tuci}.

The data are further split according to a classifier sensitive to the signal-to-background ratio. The classifier combines a number of track-related observables, including track and vertex quality, transverse momenta of \KS and \Dz candidates, helicity angles of the \KS and \Dz decays, and particle identification parameters of the \Dz final state particles. The kinematics of the tagging pion is excluded to avoid introducing possible charge asymmetry biases.
All these observables are combined in a k-nearest-neighbors (kNN) classifier~\cite{kNN}, trained using a simulated signal sample, and data from the \Dz mass sidebands for background. The resulting discriminant is used to split each sample in three categories (low, medium and high purity); the low-purity class contains very little signal and is removed. The thresholds separating the three categories are numerically optimized in a way to achieve the best combined resolution on \ACP~\cite{tuci}.

Finally, 2015--2016 data are analyzed separately from 2017--2018 data, due to different trigger conditions. 
In addition, DD candidates were not collected in 2015--2016 data taking.
In about 10\% of cases, multiple \Dstarp candidates are found in the same event. This is mainly due to \Dz candidates that are associated with multiple tagging pions. In these cases, one \Dstarp candidate is randomly selected for the analysis.

\begin{figure*}
    \centering
    \includegraphics[width=0.4\linewidth]{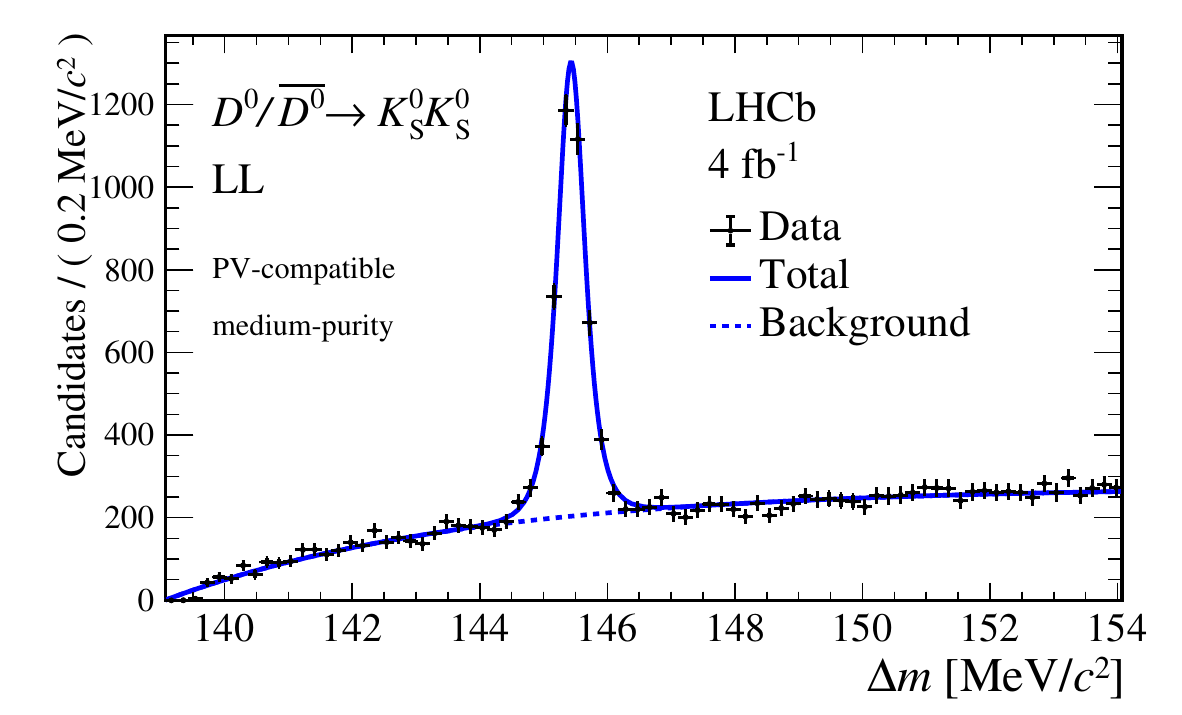}\includegraphics[width=0.4\linewidth]{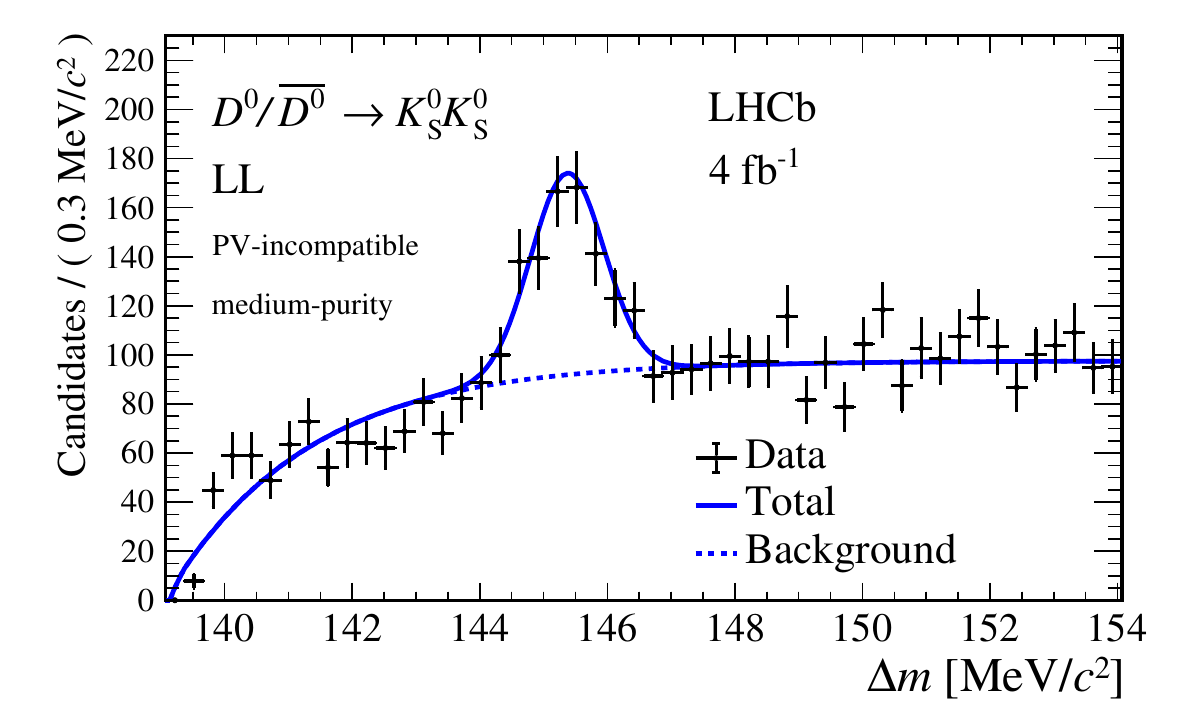}
    \includegraphics[width=0.4\linewidth]{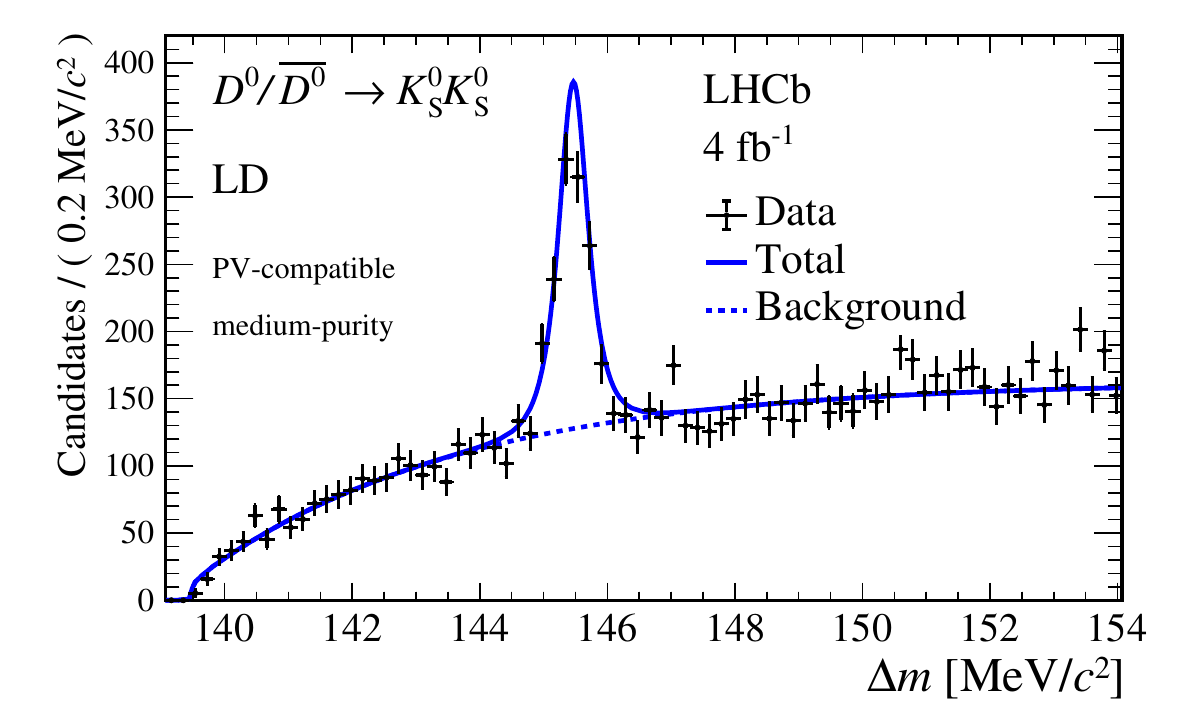}\includegraphics[width=0.4\linewidth]{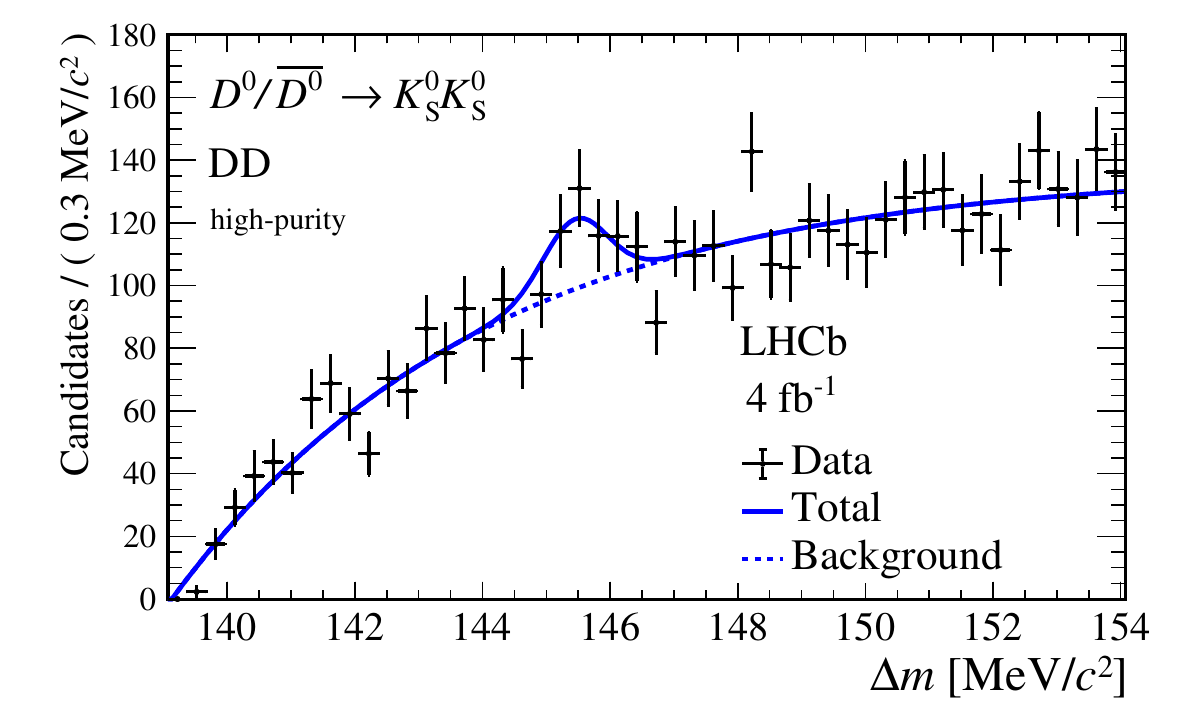}
    \caption{Distributions and fit projections of the $\Delta m$ observable for representative candidate categories (2017--2018 data). See text for the definition of purities.}
    \label{fig:yields}
\end{figure*}

Distributions of the \Delm variable for some representative subsamples are shown in Fig.~\ref{fig:yields}.
An unbinned maximum-likelihood fit is performed to the joint distribution of \Delm and the two $m(\KS)$ observables, simultaneously to candidates of both flavors to obtain the value of \ACP.
The total probability density function is parameterized by the sum of eight components: the signal component, peaking in the three observables, and seven additional components, each describing a specific background source. This includes $\Dz \to\ \KS \pip \pim$ decays, which peak in \Delm and in one $m(\KS)$ distribution but not in the other, and all possible combinations of unrelated particles. 
The peaking component in the \Delm distribution is described by a Gaussian function in PV-incompatible samples, or a Johnson $S_U$ distribution~\cite{Johnson:1949zj} in PV-compatible samples. 
The peaking component in the $m(\KS)$ distribution is described by the sum of two Gaussian functions with different widths and a common mean, for both L- and D-category \KS candidates.
The nonpeaking component in the \Delm distribution is described by an empirical threshold function, while in the $m(\KS)$ distribution it is described by a first- and second-order Chebyshev polynomial for L- and D-category \KS candidates, respectively.
In each subsample, the parameters defining the signal and background probability density functions are shared between flavors, while the normalization of each component is allowed to differ.

Each candidate participating in the fit is appropriately weighted with the aim of correcting for all spurious asymmetries, with the help of the calibration $\decay{\Dz}{\Kp \Km}$ sample, selected in a way to contain the same proportions of primary and secondary decays.  
The calibration sample is similarly split between PV-compatible and PV-incompatible categories, and is required to have $m(\Kp\Km)$ within $\pm 20\mevcc$ around the known $\Dz$ mass. In addition, only candidates in a $\pm 1.5\mevcc$ range around the \Delm peak are used. 

To compute weights, each \decay{\Dz}{\Kp \Km} subsample is classified in categories of observed charge asymmetry, by a kNN classifier based on the space of kinematic parameters of the~\Dz candidate. 
The detector and production asymmetries observed in the calibration sample are independent of the \Dz decay mode, being charge symmetric, and can be used to correct the signal sample for the same effects. This is achieved by weighting each signal candidate in the global fit by a charge-dependent factor
\begin{equation}
    w^\pm(\vec{p}_0) =  \frac{n_C^+(\vec{p}_0)+n_C^-(\vec{p}_0)}{2 n_C^\pm(\vec{p}_0)}\left[1\pm \ACP(\Kp\Km)\right],
\end{equation}
where $\vec{p}_0$ is the \Dz 3-momentum, and $n_C^\pm(\vec{p}_0)$ is the density of calibration \Dstarpm decays in the $\vec{p}_0$ space\cite{tuci}. 

\begin{figure}
    \centering
    \includegraphics[width=0.4\linewidth]{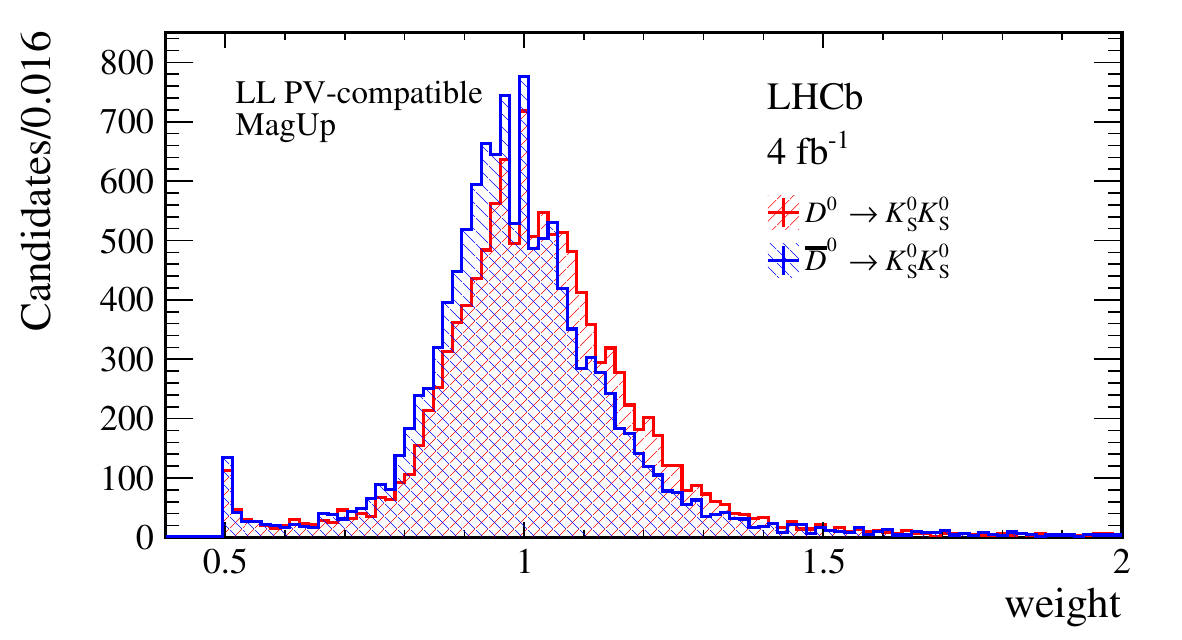}\includegraphics[width=0.4\linewidth]{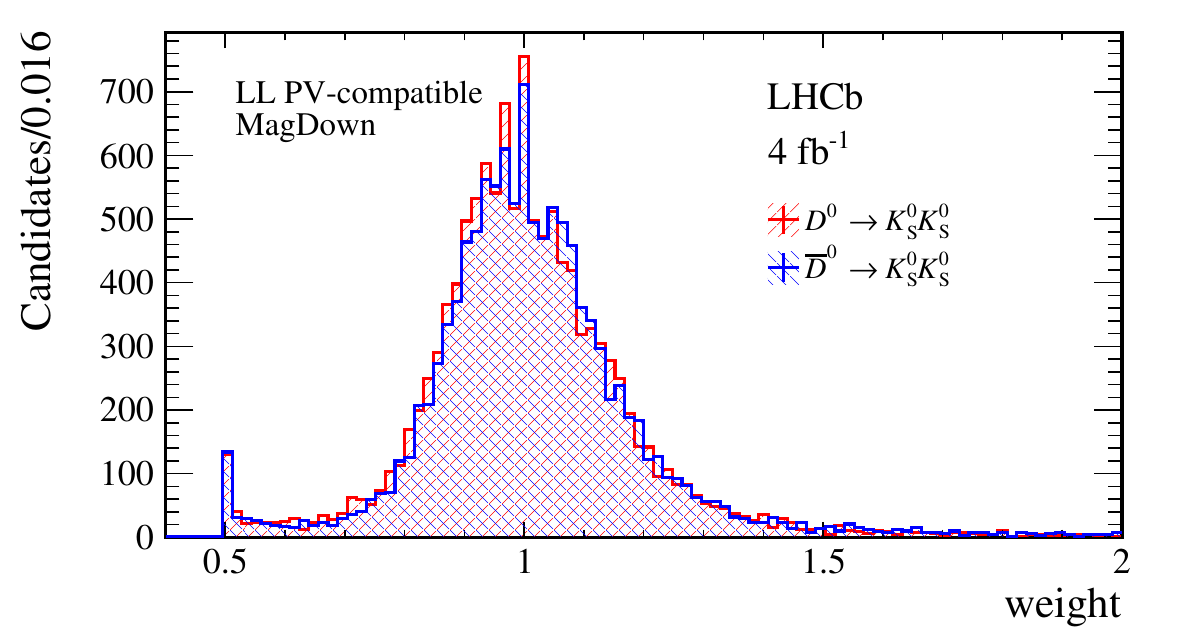}
    \caption{Distributions of weights (2017--2018 data) applied to signal candidates in the global fit to correct for detector, production, and physics of the calibration channel. Differences between \MagUp (left) and \MagDown (right) are a consequence of different \Dz and \Dzb acceptances,  and of detector asymmetries. A small fraction of entries with large weights fall outside the range of the plot. }%###}
    \label{fig:weights}
\end{figure}

To account for a possible dependence of the detection asymmetry on magnet polarity, weights are separately calculated for \MagUp and \MagDown configurations. Their distributions are shown in Fig.~\ref{fig:weights}, where their difference is clearly visible. To avoid weights affected by large uncertainties, a negligible fraction of candidates having very large weights (greater than 10) are dropped at this stage.  
The weighting procedure described above is a better alternative to the procedure of binning the \Dz kinematic space and weighting the two samples to have the same distribution. The advantage is a reduction of dimensionality to the single variable of actual relevance, that is the charge asymmetry to be corrected. This reduces the loss in statistical power that occurs when weighting bins of very different populations in the two samples under comparison. In addition, it holds exactly even when the involved asymmetries are not small. This is helpful in the case of the LHCb detector, where some kinematic regions are characterized by sizeable detection asymmetries for charged pions~\cite{LHCb-PAPER-2018-012}. This affects the tagging pion in 30\% of our events.

This weighting method has been extensively checked with both data and simulation. 
One of the checks is to apply the weighting procedure to half of the $\decay{\Dz}{\Kp \Km}$ sample, using the second half to calculate weights. 
This has the expected effect of canceling the asymmetry of the sample, that was initially highly significant, $(1.3\pm 0.1)\%$, due to a combination of physics and detector effects.

The systematic uncertainty associated to the limited size of the calibration mode is determined by a bootstrap sampling of the data, and is found to be negligible.
The uncertainty due to the presence of  residual background in the \decay{\Dz}{\Kp \Km} sample has also been evaluated and found to be a minor effect. 

Another important source of systematic uncertainty is due to the limited knowledge of the shape of the mass distributions, as the fit function used is purely empirical. Any potential bias in the procedure is evaluated by fitting the model to simulated samples of pseudoexperiments from alternative models.  
The number and type of peaking components and the behavior of the background at the kinematic threshold are varied, and the largest observed variations are taken as the systematic uncertainty, ranging between 0.4\% and 5.6\%, depending on the subsample.

A systematic uncertainty is also assessed for possible residual differences in secondary decay fractions between signal and calibration samples. The largest discrepancy between the two samples is the presence of slightly different trigger requirements on the proper decay time of the \Dz candidate. This has been corrected for by increasing the weight of candidates close to the threshold, to emulate the effect of those that have been lost. An uncertainty of this correction is conservatively assessed, that is between 0.1\% and 0.2\%.
Finally, an uncertainty of $0.15\%$ in the input value of $\ACP(\Kp\Km)$ is separately taken into account, measured by the LHCb experiment as $\ACP(\Kp\Km)=(0.04 \pm 0.12 \pm 0.10)\%$~\cite{LHCb-PAPER-2016-035}. 

The results obtained in each subsample are summarized in Table~\ref{tab:res}. The subsamples corresponding to different kNN classifier ranges are fitted simultaneously with common parameters, so they do not produce separate results. As a check of goodness of fit, a $\chisq$ has been calculated for each one-dimensional projection of the fit (Fig.~\ref{fig:yields}). All $p$-values are found to be greater than $0.2$.

\begin{table*}[tbp]
	\begin{center}
		\caption{Measurements of yields and $\ACP(\Dz\to \KS\KS)$ in individual subsamples. For asymmetries, the first uncertainty is statistical and the second is systematic.}
		\label{tab:res}
		\begin{tabular}{l|c c|c c}
			\toprule
                Sample &  \multicolumn{2}{c|}{2015 + 2016 (2 \invfb)} &  \multicolumn{2}{c}{2017 + 2018 (4 \invfb)}\\
                 & Yield & \ACP [\%] & Yield & \ACP [\%]\\
                 \midrule
                LL PV-comp. & $1388 \pm 41$ & $\phantom{(}\phantom{0}\phantom{-}0.3 \pm  \phantom{0}2.5\pm 0.6$ &  $ 4056 \pm 77$ & $\phantom{0}-4.3\pm \phantom{0}1.6 \pm 0.4 $\\
                LL PV-incomp. & $\phantom{0}178 \pm 31$ & $\phantom{}-11\phantom{.0}\pm 17\phantom{.0} \pm 2\phantom{.0}$ & $ \phantom{0}430 \pm 41$ & $\phantom{0}-3.0\pm \phantom{0}7.9 \pm 1.1 $\\
                LD PV-comp. & $\phantom{0}411 \pm 25$ & $\phantom{0}-7.2 \pm \phantom{0}5.8 \pm 1.1$  & $ 1145 \pm 49$ & $\phantom{0}-2.9\pm \phantom{0}3.8 \pm 0.7$\\
                LD PV-incomp. & $\phantom{00}58\pm 18$ & $\phantom{}-10\phantom{.0}\pm 31\phantom{.0} \pm 4\phantom{.0}$ & $ \phantom{0}349 \pm 64$ & $\phantom{0}-5\phantom{.0}\pm 17\phantom{.0} \pm 2\phantom{.0} $ \\
                DD & $-$ & $-$  & $ \phantom{00}87 \pm 28$ & $\phantom{} -35\phantom{.0}\pm 47\phantom{.0} \pm 6\phantom{.0} $\\
                \bottomrule
		\end{tabular}
	\end{center}
\end{table*}

All partial results in Table~\ref{tab:res} are statistically compatible with each other. The weighted average of all measurements using 2015--2016 data, is
$$\ACP(\KS\KS)= (-1.1 \pm 2.3\pm 0.5  \pm 0.2) \%,$$
where the first uncertainty is statistical; the second is systematic, obtained by taking the individual contributions as uncorrelated; and the third is from the uncertainty on \mbox{$\ACP(\Kp\Km)$}. This result is compatible with the previously published value based on the same data sample~\cite{LHCb-PAPER-2018-012}, but has a better precision by about $30$\%, corresponding to an effective doubling of the yields. The sensitivity increase is due to the combined effect of several improvements in the analysis, most notably the new weighting technique, the inclusion of secondary decays, an appropriate categorization of the sample, and the multidimensional likelihood fit. 

The asymmetry for the 2017--2018 data is measured to be
$$\ACP(\KS\KS)=(-4.0 \pm 1.5 \pm 0.3 \pm 0.2)\% .$$
Treating the systematic uncertainties as uncorrelated between the data taking periods, except for the shape of the fit functions that is considered to be fully correlated, the asymmetry combining the results from the full Run 2 data sample is obtained as

$$\ACP(\KS\KS) =(-3.1 \pm 1.2 \pm 0.4 \pm 0.2)\% .$$
This measurement supersedes the previous LHCb result~\cite{LHCb-PAPER-2018-012} and is in agreement with all previous determinations~\cite{Bonvicini:2000qm,Dash:2017heu,LHCb-PAPER-2015-030}.  It is the most precise measurement of this quantity to date, and it is compatible with no \CP asymmetry at the level of $2.4$ standard deviations.

\section*{Acknowledgements}
%
% These Acknowledgements valid from 3-May-2019
%
\noindent We express our gratitude to our colleagues in the CERN
accelerator departments for the excellent performance of the LHC. We
thank the technical and administrative staff at the LHCb
institutes.
We acknowledge support from CERN and from the national agencies:
CAPES, CNPq, FAPERJ and FINEP (Brazil); 
MOST and NSFC (China); 
CNRS/IN2P3 (France); 
BMBF, DFG and MPG (Germany); 
INFN (Italy); 
NWO (Netherlands); 
MNiSW and NCN (Poland); 
MEN/IFA (Romania); 
MSHE (Russia); 
MICINN (Spain); 
SNSF and SER (Switzerland); 
NASU (Ukraine); 
STFC (United Kingdom); 
DOE NP and NSF (USA).
We acknowledge the computing resources that are provided by CERN, IN2P3
(France), KIT and DESY (Germany), INFN (Italy), SURF (Netherlands),
PIC (Spain), GridPP (United Kingdom), RRCKI and Yandex
LLC (Russia), CSCS (Switzerland), IFIN-HH (Romania), CBPF (Brazil),
PL-GRID (Poland) and NERSC (USA).
We are indebted to the communities behind the multiple open-source
software packages on which we depend.
Individual groups or members have received support from
ARC and ARDC (Australia);
AvH Foundation (Germany);
EPLANET, Marie Sk\l{}odowska-Curie Actions and ERC (European Union);
A*MIDEX, ANR, IPhU and Labex P2IO, and R\'{e}gion Auvergne-Rh\^{o}ne-Alpes (France);
Key Research Program of Frontier Sciences of CAS, CAS PIFI, CAS CCEPP, 
Fundamental Research Funds for the Central Universities, 
and Sci. \& Tech. Program of Guangzhou (China);
%Key Research Program of Frontier Sciences of CAS, CAS PIFI,
%Thousand Talents Program, and Sci. \& Tech. Program of Guangzhou (China);
RFBR, RSF and Yandex LLC (Russia);
GVA, XuntaGal and GENCAT (Spain);
the Leverhulme Trust, the Royal Society
 and UKRI (United Kingdom).

\addcontentsline{toc}{section}{References}
%\setboolean{inbibliography}{true}
\bibliographystyle{LHCb}
\bibliography{main,standard,LHCb-PAPER,LHCb-CONF,LHCb-DP,LHCb-TDR}

\newpage
% LHCb collaboration author list
% Data extracted on September 15th, 2021 at 9:12pm for paper reference LHCb-PAPER-2020-047
\centerline
{\large\bf LHCb collaboration}
\begin
{flushleft}
\small
R.~Aaij$^{32}$,
C.~Abell{\'a}n~Beteta$^{50}$,
T.~Ackernley$^{60}$,
B.~Adeva$^{46}$,
M.~Adinolfi$^{54}$,
H.~Afsharnia$^{9}$,
C.A.~Aidala$^{85}$,
S.~Aiola$^{25}$,
Z.~Ajaltouni$^{9}$,
S.~Akar$^{65}$
% ...

{\footnotesize \it

% ...
$^{9}$Universit{\'e} Clermont Auvergne, CNRS/IN2P3, LPC, Clermont-Ferrand, France\\
% ...
$^{25}$INFN Sezione di Milano, Milano, Italy\\
% ...
$^{32}$Nikhef National Institute for Subatomic Physics, Amsterdam, Netherlands\\
% ...
$^{46}$Instituto Galego de F{\'\i}sica de Altas Enerx{\'\i}as (IGFAE), Universidade de Santiago de Compostela, Santiago de Compostela, Spain\\
% ...
$^{50}$Physik-Institut, Universit{\"a}t Z{\"u}rich, Z{\"u}rich, Switzerland\\
% ...
$^{54}$H.H. Wills Physics Laboratory, University of Bristol, Bristol, United Kingdom\\
% ...
$^{60}$Oliver Lodge Laboratory, University of Liverpool, Liverpool, United Kingdom\\
% ...
$^{65}$University of Cincinnati, Cincinnati, OH, United States\\
% ...
$^{68}$Syracuse University, Syracuse, NY, United States\\
% ...
$^{85}$University of Michigan, Ann Arbor, United States, associated to $^{68}$\\
\bigskip
% ...
\medskip
}
\end{flushleft}

\end{document}